\documentclass[journal]{IEEEtran}
\usepackage{comment}
\usepackage{algorithm}
\usepackage{algorithmic}
\usepackage{bm}
\usepackage{array}
\usepackage{graphicx}
\usepackage{booktabs}
\usepackage{amsfonts,amssymb,amsmath}
\usepackage{subfigure}

\begin{document}
%
\title{Network Coding Tree Algorithm for Multiple Access System}
%
%
%

\author{\authorblockN{Zhengchuan Chen\authorrefmark{1}\thanks{An extension of this work with theoretical analysis on the throughput and delay of the network coding tree algorithm has been submitted.}, Ke Xiong\authorrefmark{2}, Pingyi Fan\authorrefmark{1} and Chen Chen\authorrefmark{1}\\
\authorblockA{\authorrefmark{1}Tsinghua National Laboratory for Information
Science and Technology (TNList),\\
and Department of Electronic Engineering,
Tsinghua University, Beijing,
China.}\\
\authorblockA{\authorrefmark{2}School of Computer and Information Technology, Beijing Jiaotong University, Beijing, China.}\\
\authorrefmark{1}\{chenzc10@mails.,fpy@,c-c10@mails.\}@tsinghua.edu.cn, \authorrefmark{2}kxiong@bjtu.edu.cn } }

\maketitle

\begin{abstract}
Network coding is famous for significantly improving the throughput of networks. The successful decoding of the network coded data relies
on some side information of the original data. In that framework,
independent data flows are usually first decoded and then network
coded by relay nodes. If appropriate signal design is adopted, physical layer network coding is a natural way in wireless
networks. In this work, a network coding tree algorithm which
enhances the efficiency of the multiple access system (MAS) is
presented. For MAS, existing works tried to avoid the collisions while
collisions happen frequently under heavy load. By introducing
network coding to MAS, our proposed algorithm achieves a better
performance of throughput and delay. When multiple users transmit
signal in a time slot, the mexed signals are saved and used to jointly
decode the collided frames after some component frames of the
network coded frame are received. Splitting tree structure is
extended to the new algorithm for collision solving. The throughput
of the system and average delay of frames are presented in a
recursive way. Besides, extensive simulations show that network
coding tree algorithm enhances the system throughput and decreases the average frame delay compared with other
algorithms. Hence, it improves the system performance.
\end{abstract}

%

\section{Introduction}
The medium access control (MAC) protocols play an
important part in modern communication networks such as
Internet and wireless local area network (WLAN). Different protocols share the multiple access
medium among users in different way. They are usually divided into
two types. The first type is called a reservation protocol, in which
users make reservations in the first stage in order to determine the transmission scheduling in the next stage. The second type is called a
direct transmission protocol, in which users send their frames
following certain rules based on the channel feedback.
The feedback often indicates that zero, one or two and more frames
(collision state) are transmitted in the last slot \cite{Tsybakov}.
The core of both types of protocol rests with the collision solving
algorithms which dictate how to solve the collision situation.

Network coding has brought underlying change to the communication
world \cite{Ahlswede} \cite{Ho}. It was first proposed to achieve
the capacity of the multicast network. Later, the extended version
of this discovery-the physical layer network coding (PHY NC) was
applied to many practical situations such as two-way relay networks \cite{Zhang}, vehicular communications \cite{Qing}
and butterfly networks \cite{Chen}. PHY NC shows great potential in
improving the performance of these networks \cite{Liew}.

The typical examples of collision solving algorithm include the
slotted ALOHA algorithm and binary tree algorithm. It is well known
that the slotted ALOHA algorithm has a limit throughput of
$\frac{1}{e}$ and also requires some kinds of stabilizing method
\cite{Abramson1} \cite{Abramson2}. In order to avoid these defects,
the binary tree algorithm is proposed later \cite{Capetanakis}. It
is proved to be able to maintain stability and increase the achievable
throughput \cite{Capetanakis2}. Several modifications are made to
the binary tree algorithm to achieve a larger throughput. In both
binary tree algorithm and its extended version, the access point (AP)
of the multiple access system (MAS) drops the sum signal of the
collided frames. It is often assumed that combination of the
collision frames is useless because the frames interfere with each
other. The thought brought by PHY NC, which makes use of the
collision frames, can be used to improve the performance of
classical collision solving algorithms. After successful decoding
some raw component of the collided frames, one of the original frame
of the collided frames can be decoded successfully by regarding the
received raw component frames as side information. It is shown that
both the throughput and delay\footnote{It is noted that delay is another important performance metric in wireless communications, \cite{Zhou} and \cite{Wang} and reference therein include examples of delay-considered optimal protocol design for other wireless scenarios} of slotted ALOHA algorithm are
improved combining with PHY NC \cite{Cocco} due to that multiple
received copies introduce a diversity gain. However, in \cite{Cocco}
the authors assumed that the AP can know which terminals collide.
This assumption is too strong to implement the slotted ALOHA
algorithm into MAS. In \cite{Ni}, the authors considered to combine
the network coding strategy with ALOHA algorithm while simplifies
the physical channel model as Galois Field addition and
complication. Because the ALOHA algorithm dominates
the main performance essentially, the throughput of the algorithm proposed in
\cite{Ni} always holds a increase-decrease characteristic in terms
of the arrival rate. As binary tree algorithm is more efficient than
ALOHA algorithm when system is under heavy load \cite{Gallager}, we combine PHY NC
with the binary tree algorithm to achieve even greater performance
in this paper.

The rest of the paper is organized as follows. In Section II, we
introduce the system model and claim some necessary preliminaries. In Section III, we
propose a network coding tree algorithm (NCTA). In Section IV, we compare the throughput
and average delay of NCTA to those of binary tree algorithm (BTA)
and ALOHA system by simulation. Conclusions are drawn in Section V.

\section{System Model and Preliminaries}
\subsection{system model}
Let us consider a MAS with finite $m$ users and only one AP. Assume
the system runs with slots and the synchronization among the users and
the AP is guaranteed. In each slot, one user can transmit a frame if
other users keep silent. If two or more users transmit their
frame in the same slot, signals are collided and the AP cannot
decode any frame. At the end of each slot, we assume that users are able to get a feedback
via listening the channel or from the AP. For instance, in a bus system, AP is nothing
different from a user from a physical layer viewpoint. The
assumption that users can obtain the channel feedback via listening
is reasonable. In a WLAN system,
limited bandwidth can provide necessary
channel feedback from AP to users. 
In this paper, we further consider that each user can know from the feedback that
a collision, or a successful transmission, or nothing happens
in the last slot. For clarity, let us use a uniform function to represent the feedback and neglect the
physical layer system setup. Denote the signal received at AP
by $x$. We define
\begin{equation}
f(x)=\begin{cases} 0,~\text{$x$ made up of noise}\\
1,~\text{$x$ includes one frame}\\
e,~\text{otherwise}
\end{cases}
\end{equation}
where $e$ is used to represent the state when collision is occurring.

The frame arrivals are assumed to be independent Poisson
flows. Let us denote the sum arrival rate of the system by
$\lambda$. Accordingly, the arrival rate at each user is $\lambda/m$
while frames are poisson arrived. Usually, in MAS, if the frame
collides with others, it will be retransmitted in later slots until
it is received successfully. We will extend this rule and reduce the
retransmission times by deploying signal cancellation which realizes
network decoding. In most of the existing works, users were assumed to have an infinite long
buffer. To focus on planting network coding into MAS and investigating the throughput-delay performance gain induced by
network coding idea,
we try to simplify the system model in this
paper by considering that the buffer length is one. That is,
the buffer is modeled as a binary state machine\footnote{This assumption is used to simplify the theoretical analysis of the throughput and average delay, which is strongly related to the probability of empty buffer at each node. We will investigate the general case with finite buffer length $L$ in another paper.}. If there is a frame
at the user, it is in `active' state, otherwise it is in `inactive' state.

To compare the performance with other algorithms, throughput and
average delay are regarded as the two main performance metrics of interest in
this work. On one hand, throughput stands for the synthesized system
efficiency and it is the performance of interest in most cases. On
the other hand, under the throughput metric, the average delay play
an important role in describing the quality of service (QoS) provided by a system
\cite{Kleinrock}. For clarity, we use $\Phi$ and $D$ to represent them in our analysis,
respectively.

To analyze the MAS embedded with PHY NC, we may first review some
classical collision solving algorithms.

\subsection{Time Division Multiplexing}
In time division multiplexing (TDM), time slots are assigned to
users. Each user has to wait for his own time slot to transmit
frames. There is no frame collision under this scheme. System is
expected to transmit a frame in each time slot under heavy load.
Hence, TDM achieves very well performance with a large arrival rate. The highest
throughput can even approach $1$. Neverthless, for light load, many time
slots are idle and wasted since there is no user to transmit. Moreover, the fixed
transmission order of TDM also induces a large average delay for all arrival
rate $\lambda\in[0,1)$. The average delay, $D_{TDM}$, of TDM
is about to $\frac{m}{2}$ \cite{Gallager}.

\subsection{Slotted ALOHA Protocol}
In order to diminish the delay of the MAS, ALOHA scheme was presented which allows
users to compete for accessing the channel. If frames collide in a
time slot, each of them will be retransmitted after a
random delay with a probability $p$. In the light load case, less collisions occur so that the large delay endured in TDM scheme is significantly
alleviated. Nevertheless, even in the moderate load case, the frequent
collision may push the delay up and the system then goes to an
instable status along with increasing data arrival rate dramatically. In low arrival rate case, ALOHA has better performance. Moreover, as it can be performed
in a distributed manner, it is widely employed in lots of network scenarios.

\subsection{Binary Tree Algorithm}
Instead of letting all the users with data competing for the
channel in each time slot, BTA only allows part of the users to
retransmit their data frames in the following slots to diminish the
collision probability\cite{Gallager}. Once a collision occurs, the users are randomly split into subsets. For each subset, users retransmit their
frames in a future slot. Dividing is along with collision. If no collision happens for a time slot, the frame transmitted by some user in
the subset will be successfully decoded or none of the users has
frames to transmit\cite{Gallager}. BTA bridges ALOHA and TDM in terms of the
throughput and average delay performance. In BTA,
mixed signals of collided frames are dropped by AP. To utilize the
information contained in collided signals, we propose the NCTA in the following.

\section{Network Coding Tree Algorithm}
NCTA utilizes the mixed signal of the collided frames as a physical layer
network coded signal. Again, dividing follows collision, either
in a random or a fixed manner. The successful frame receiving of a
subset helps it decode frames of sibling subsets via canceling signal of
the successfully decoded frames from the collided mixed signals. As only the status is fed back by AP, including free time slot, collision and successful
transmission, after canceling
the decoded signal from the mixed signals, it is feasible for AP to accurately feed back the status of the rest of the signals by observing the mixed signals. So, the AP may know no frame, one frame, or two or more data frames
included in the rest of the mixed signals.

To analyze $\Phi$ and $D$, we take two types of system waiting time
into consideration. In the first type, the buffer observes only one
time slot, that is, the buffer only keeps the frames arrived in the
last slot. We call this system setup as \emph{System Type I}. In the
other case, waiting time is defined as the number of time slots used for
solving the last collision corresponding to the root node of the
algorithm tree. This system setup is called as \emph{System Type II}. Let us denote the system waiting time by $W$.

To describe the algorithm clearly, we present the pseudo code as
Algorithm \ref{NCTA}.

\begin{algorithm}[htb]
\caption{Network Coding Tree Algorithm} \label{NCTA}
\begin{algorithmic}[1]
\REQUIRE ~~\\
Set waiting time $W=1$; Clear all the user buffers; Initialize two
new stacks $S_1$ and $S_2$; Denote all the users by $A$
\ENSURE ~~\\
\WHILE {$A\neq\emptyset$} \STATE All the users with non-empty buffer
transmit; Denote the sum signal by $y$; \IF{$f(y)\neq e$} \STATE Set
$W=1$; \STATE Break; \ELSE \STATE $S_1\leftarrow A$; \STATE
$S_2\leftarrow y$; \ENDIF \WHILE {$S_1$ is not empty} \STATE
$A'\leftarrow$ Pop an element from $S_1$; \STATE $y'\leftarrow$ Pop
an element from $S_2$; \STATE Uniformly divide $A'$ into two subsets
$A_1$ and $A_2$; \STATE All the users in $A_1$ transmit frame in
next time slot; \STATE Denote the sum signals received at AP by $y_1$.
\IF{$f(y_1)\neq e$ AND $f(y'-y_1)=e$} \STATE $S_1\leftarrow A_2$;\STATE
$S_2\leftarrow y'-y_1$; \ELSE\IF{$f(y_1)=e$ AND $f(y'-y_1)\neq e$} \STATE
$S_1\leftarrow A_1$; \STATE $S_2\leftarrow y_1$;
\ELSE\IF{$f(y_1)=e$ AND $f(y'-y_1)=e$} \STATE $S_1\leftarrow A_2$; \STATE
$S_1\leftarrow A_1$; \STATE $S_2\leftarrow y'-y_1$\STATE
$S_2\leftarrow y_1$ \ENDIF \ENDIF \ENDIF \ENDWHILE \STATE Set $W$
according to the time slots used for collision solving.\ENDWHILE
\end{algorithmic}
\end{algorithm}

Note that $y'-y_1$ can be regarded as the signal transmitted by nodes in $A_2$, polluted by noisy of transmission corresponding to $A'$ and that of transmission corresponding to $A$.

\section{Simulations}
We present the throughput and average delay of the new proposed network coding tree algorithm and other known algorithms with respect to the arrival rate $\lambda$ for two types of system in this section. To compare NCTA, BTA with ALOHA, we modify the ALOHA system to
guarantee that the system conditions for these algorithms are the
same. We state the modified ALOHA system as follows.

At the beginning, the system is with unblocked state. If there are
frames arriving at some users, then those users transmit frame. If no
collision occurs, the system maintains unblocked state, otherwise it goes into
blocked state. In blocked state, all the users with frames transmit
in each time slot with probability $p$ until the frame is
successfully received by AP. After that, system returns to unblocked
state. In the modified ALOHA algorithm, if there are $n$ users with
frame, then $p$ is set to $\frac{1}{n+1}$. Because it can maximize
the probability of successful transmission, namely, $C_n^1p(1-p)^n$
in next time slot. In the rest of this paper, we use ALOHA to
represent the modified ALOHA system without ambiguity.

\begin{figure}[t]
\centering
\includegraphics[width=0.5\textwidth]{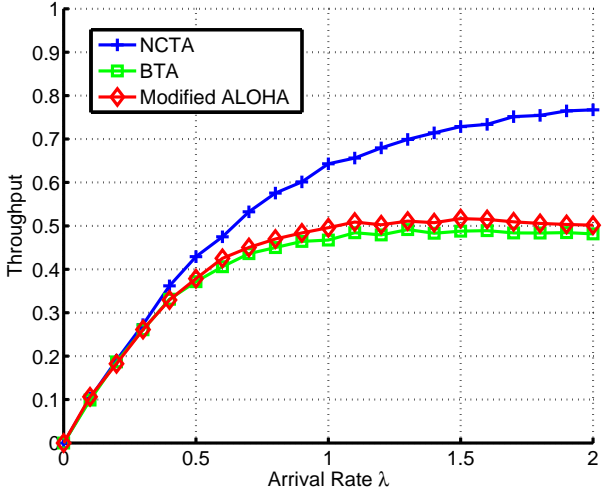}
\caption{The throughput of network coding tree algorithm (NCTA), binary tree algorithm (BTA) and modified ALOHA system for system type I.} \label{Th_T1}
\end{figure}

First, for \emph{System Type I}, we compare the throughput of NCTA, BTA and
ALOHA. We consider $M=8$ and the arrival rate of the system,
$\lambda\in[0,2]$. For every point, it is the average of 5000
simulations. As illustrated in Fig. \ref{Th_T1}, for all the arrival rate
$\lambda\in[0,2]$, the throughput of NCTA is clearly greater than
that of BTA and ALOHA. Among them, the throughput of NCTA ascends in
terms of $\lambda$ and keeps stable around 0.75 while that of NC
and ALOHA, keeps 0.5. That is, the throughput is increased by 0.25.
\begin{figure}[t]
\centering
\includegraphics[width=0.5\textwidth]{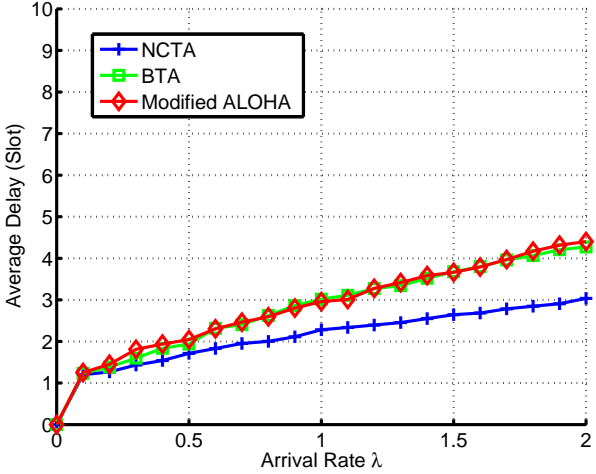}
\caption{The average delay of network coding tree algorithm (NCTA), binary tree algorithm (BTA) and modified ALOHA system for system type I.} \label{Delay_T1}
\end{figure}
Average delays of different algorithms corresponding to this setup are depicted in Fig. \ref{Delay_T1}.
Again, for all the arrival rate, the average delay of NCTA is less than that
of BTA and ALOHA while that of BTA and ALOHA are
roughly the same with each other.

In summary, under \emph{System Type I}, both throughput and average delay
of NCTA are better than those of BTA and ALOHA.

\begin{figure}[t]
\centering
\includegraphics[width=0.5\textwidth]{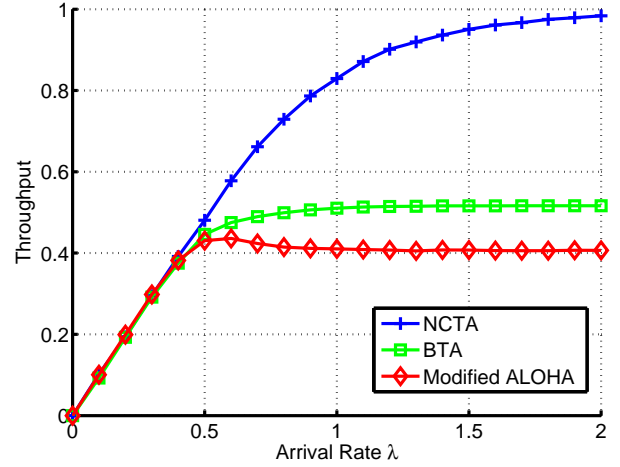}
\caption{The throughput of network coding tree algorithm (NCTA), binary tree algorithm (BTA) and modified ALOHA system for system type II.} \label{Th_T2}
\end{figure}

\begin{figure}[t]
\centering
\includegraphics[width=0.5\textwidth]{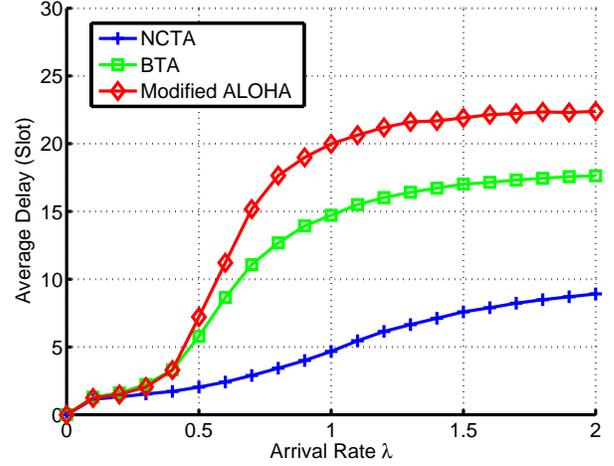}
\caption{The average delay of network coding tree algorithm (NCTA), binary tree algorithm (BTA) and modified ALOHA system for system type II.} \label{Delay_T2}
\end{figure}

For \emph{System Type II}, to keep consistent with that of \emph{System Type I},
we assume $M=8$ and $\lambda\in[0,2]$ again. For every point, we
average out 5000 simulations. The throughput and average delay of
the three algorithms are shown in Fig. \ref{Th_T2} and Fig. \ref{Delay_T2}. In Fig. \ref{Th_T2}, we can
see the maximal throughput of the modified ALOHA system is 0.45 and
keeps 0.4 when throughput is stable, which is greater than that of
the slotted ALOHA, $\frac{1}{e}\approx0.37$. The throughput of BTA,
is about 0.5 while that of NCTA, is significantly approaching 1 when
$\lambda$ is increasing. Before it achieves the maximum, the
throughput of the three algorithms are increasing linearly with the
arrival rate. After that, the throughput of ALOHA drops slightly.
This shows that NCTA and BTA are more stable than ALOHA.


Comparing Fig. \ref{Th_T1} with Fig. \ref{Th_T2}, it is not hard to see that for \emph{System Type II}, the throughput of ALOHA decreases while that of BTA improves a little and that of NCTA increases significantly. Under
\emph{System Type II}, the system waiting time is longer, hence, the
probability of collision becomes larger. This implies that ALOHA
system is inefficient to solve collision while NCTA is good at
solving collisions with many frames.

Fig. \ref{Delay_T2} shows the average delay of three algorithms under \emph{System Type II}. Along with the increasing of $\lambda$, all the average delay
curves keep ascending. For all the arrival rate, the average delay
of NCTA is always less than those of BTA and ALOHA. Besides, when
$\lambda=0.4$, the slope of the delay curves increase dramatically.
The slope roughly reflects the stability of the algorithm. From Fig. \ref{Delay_T2}, we can see that the slope of ALOHA's average delay is the largest one followed by that of BTA and that of NCTA is the smallest.
Comparing with the delay curves of \emph{System Type I}, systems bear a
larger delay for all the three algorithms. This is induced by the
increased quantity of collided frames, hence, the increased time
slots used for collision solving.

\begin{figure}[t]
\centering
\includegraphics[width=0.5\textwidth]{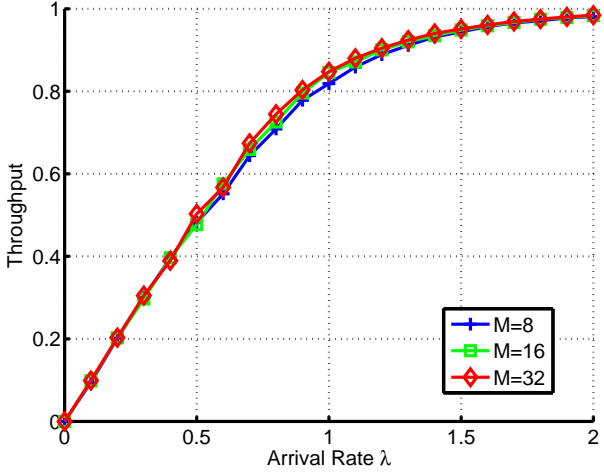}
\caption{The throughput of network coding tree algorithm for system type II. $M$ represents the system user number.} \label{Th_M}
\end{figure}
\begin{figure}[t]
\centering
\includegraphics[width=0.5\textwidth]{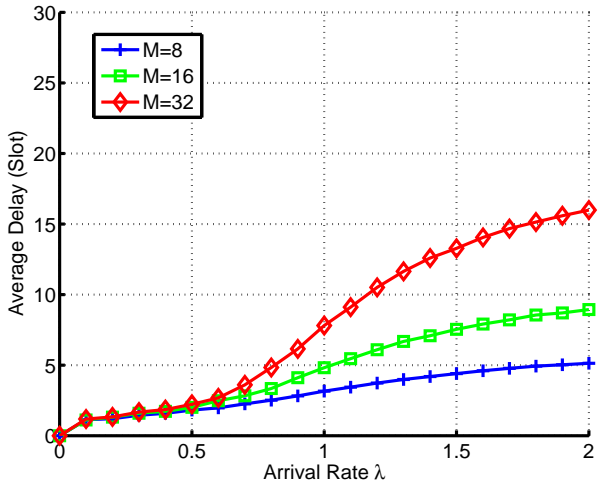}
\caption{The average delay of network coding tree algorithm for system type II. $M$ represents the system user number.} \label{Delay_M}
\end{figure}

Next, we consider the impact of system user quantity under system
type II. In Fig. \ref{Th_M} and Fig. \ref{Delay_M}, the throughput and the average delay of
NCTA for $m=8$, $m=16$, $m=32$ are illustrated. In Fig. \ref{Th_M}, we can
see that for given arrival rate, the throughput of the system is
increased slightly along with the increasing user number. It is
predicted that when $m\rightarrow+\inf$, the throughput will stable
around a fixed value. Moreover, the income induced by increasing the
user number is marginal. In Fig. \ref{Delay_M}, it is not difficult to see that
along with the increasing user number $m$, the average delay of the
frames is blowing up. The slope of the delay curve is ascending
dramatically. The reason is that when
system has more users, the collided frames are more than that of a
system with less users. there are many frames involved in a
collision, therefore, the time taken to solve the collision will be
longer. In average, each frame has to wait long time to be decoded
successfully by AP. According to Fig. \ref{Th_M} and Fig. \ref{Delay_M}, when user number falls down, the
throughput decreases very slightly while the
average delay of the frames, drops down drastically.

\begin{figure}[!ht]
\centering
\includegraphics[width=0.5\textwidth]{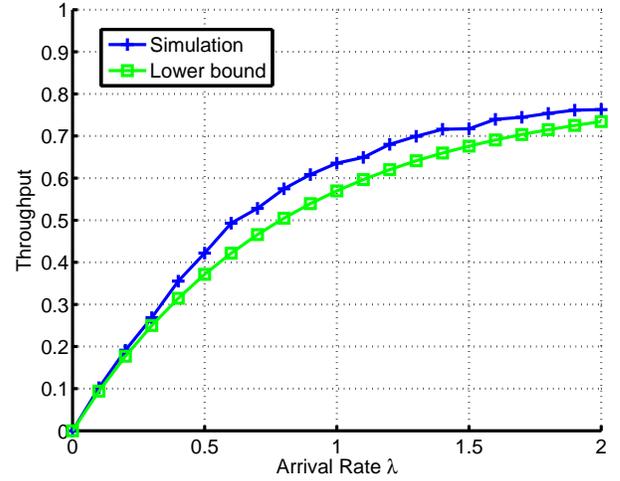}
\caption{The simulation result and lower bound for the throughput of network coding tree algorithm, system type II. } \label{Th_LB}
\end{figure}

\section{Conclusion}
In this work, we introduced network coding into multiple access
system and propose a network coding tree algorithm. The implicit
form of the throughput and average delay of the system operated with
the new algorithm were given in an iterative way. To show the
performance of the algorithm, we compare the throughput and average
delay induced by the new algorithm with that of modified
ALOHA scheme and tree algorithm via simulation. The results showed
that when operated with network coding tree algorithm, multiple
access system can achieve a larger throughput while the frames bear
shorter average delay. Besides, when system has more users, the throughput
increases slightly while the average delay ascends drastically.

\ifCLASSOPTIONcaptionsoff
  \newpage
\fi



%

\end{document}